\documentclass[12pt,preprint]{emulateapj}

\slugcomment{AJ accepted}

\shorttitle{LSB Galaxies and their GCs}
\shortauthors{Villegas et al.~}

\begin{document}

\title{Normal Globular Cluster Systems in Massive Low Surface Brightness Galaxies\altaffilmark{1}}

\author{Daniela Villegas\altaffilmark{2,3}, Markus Kissler-Patig\altaffilmark{2}, Andr\'es Jord\'an\altaffilmark{4,2,3,5}, Paul Goudfrooij\altaffilmark{6} and Martin Zwaan\altaffilmark{2}}

\altaffiltext{1}{Based on observations with the NASA/ESA Hubble Space Telescope, obtained at the Space Telescope Science Institute (STScI), which is operated by the Association of Universities for Research in Astronomy, Inc., under NASA contract \mbox{NAS 5-26555}.}
\altaffiltext{2}{European Southern Observatory, Karl-Schwarzschild-Strasse 2, 
85748 Garching bei M\"unchen, Germany}
\altaffiltext{3}{ Departmento de Astronom\'ia \& Astrof\'isica, 
P. Universidad Cat\'olica de Chile, Casilla 306, Santiago 22, Chile}
\altaffiltext{4}{Harvard-Smithsonian Center for Astrophysics,
Cambridge, MA 02138, USA}
\altaffiltext{5}{Clay fellow}
\altaffiltext{6}{Space Telescope Science Institute, 3700 San Martin Drive, 
Baltimore, MD 21218, USA}

\begin{abstract}
We present the results of a study of the globular cluster systems of 6 massive spiral galaxies, originally cataloged as low surface brightness galaxies but here shown to span a wide range of central surface brightness values, including two intermediate to low surface brightness galaxies. We used the Advanced Camera for Surveys on board HST to obtain photometry in the F475W and F775W bands and select sources with photometric and morphological properties consistent with those of globular clusters. A total of 206 candidates were identified in our target galaxies. From a direct comparison with the Galactic globular cluster system we derive specific frequency values for each galaxy that are in the expected range for late-type galaxies. We show that the globular cluster candidates in all galaxies have properties consistent with globular cluster systems of previously studied galaxies in terms of luminosity, sizes and color. We establish the presence of globular clusters in the two intermediate to low surface brightness galaxies in our sample and show that their properties do not have any significant deviation from the behavior observed in the other sample galaxies. Our results are broadly consistent with a scenario in which low surface brightness galaxies follow roughly the same evolutionary history as normal (i.e.~high surface) brightness galaxies except at a much lower rate, but require the presence of an initial period of star formation intense enough to allow the formation of massive star clusters. 
\end{abstract}


\section{Introduction}

Our understanding of the distribution of the central surface brightness of late-type galaxies has evolved significantly in the last decades. Late-type galaxies were originally thought to show a limiting central surface brightness value in the B band of $\mu_{0}$=21.65 mag/arcsec$^2$ (the so-called Freeman value; Freeman 1970), but Disney's (1976) realization of the severe selection effects that the sky brightness introduces in the observations led to the current knowledge of the existence of significant numbers of disk galaxies with much lower values of central surface brightness. McGaugh, Schombert \& Bothun (1995) showed, using a selection-effects-corrected surface brightness distribution, that the number of galaxies faintwards of the Freeman value remains approximately constant. This implies that the spatial density of the objects known as ``low surface brightness galaxies'' ($\mu_{0}\gtrsim$ 23 mag/arsec$^2$) is $\sim$10$^5$ times higher than what is expected based on a distribution based on Freeman's value.

Low surface brightness (LSB) galaxies encompass many of the extremes in galaxy properties. The vast majority are late-type disk galaxies with high gas mass fractions (Schombert et al.~1992), the surface density of this gas being well below the critical threshold density for star formation (van der Hulst et al.~1993). Therefore, as it has been confirmed from H$\alpha$ studies (Schombert 1990), LSB galaxies have very low star formation rates. They also appear to be more isolated than normal high surface brightness (HSB) galaxies (Bothun et al.~1993, Mo et al.~1994, Rosenbaum \& Bomans 2004) and typically show low metal content (de Blok \& van der Hulst 1998). All this observational evidence seems to indicate that LSB galaxies are relatively unevolved and quiescent objects, and may therefore provide an insight into the evolution of galaxies in unperturbed environments (de Blok et al.~1995).

Several evolutionary scenarios have been proposed to account for the properties of LSB galaxies. Based on their observed blue colors, de Blok et al.~(1995) conclude that LSB galaxies cannot be the faded remnants of normal disk galaxies. More recently, van den Hoek et al.~(2000) have shown that LSB galaxies roughly follow the same evolutionary history as HSB galaxies, except at a much lower rate. Using near-infrared observations of a sample of 88 LSB galaxies, Galaz et al.~(2002) show that a high fraction of LSBs have a well-developed old stellar population and that older LSBs are more frequent than optical data suggests. These observations are consistent with the scenario of Dalcanton, Spergel \& Summers (1997), in which LSB galaxies are formed within a hierarchical formation scenario from low mass and/or high angular momentum proto galaxies which naturally form low baryonic surface density disks (see also Jimenez et al.~1998). 
On the other hand, Zackrisson, Bergvall \& Ostlin (2005) argue using observations of extremely blue LSB galaxies that their properties are inconsistent with constant star formation rates over cosmological timescales and that current observations cannot rule out the alternative possibility that these objects formed as recently as 1--2 Gyr ago. Therefore, some LSB galaxies could have formed very recently.

Globular cluster (GC) systems around galaxies are useful tracers to constrain the formation and evolution of their hosts (e.g., West et al.~2004). They trace all the major epochs of star formation in a galaxy and their luminosities and colors contain information about the chemical enrichment environment and epoch in which they formed. Hence, the identification and study of the properties of GCs in LSB galaxies will allow to obtain a physical understanding regarding the differences and similarities between the formation and evolution processes of LSB and normal galaxies by comparing the properties of their GC systems. 

Here we present the results of a study of GC systems in a group of 6 galaxies, originally selected as LSB that turned out to span a range in central surface brightness, from normal to low surface brightness ones. This is the first study that aims to identify such objects in {\it massive} LSB galaxies. A previous study by Sharina et al.~(2005) identified a number of cluster candidates in LSB galaxies using WFPC2 images, but their sample was constrained to dwarf galaxies. Our aim in this work is to probe for the presence of globular clusters in \textit{massive} LSB galaxies and to compare their general properties to those observed in normal HSB galaxies. In particular, we aim to study the properties of objects consistent with being {\it old} GCs and which thus probe the early-stages in the formation of LSB galaxies. The mere presence of old GCs in these galaxies directly implies that there was an early period of star formation intense enough to produce massive star clusters.

This paper is organized as follows. The observations and data reduction procedures are described in \S\ref{sec:OBS&DATA}. A study of the light profiles of the target galaxies aimed to obtain an improved classification of the galaxies in terms of their central surface brightness is presented in \S\ref{sec:SBP}, while \S\ref{sec:GCsel} describes the selection of bona-fide GC candidates and a set of control-fields for each galaxy. In \S\ref{sec:TotalGC} we use our observations to estimate the total number of GCs in our target galaxies and we discuss the properties of the detected GC systems in \S\ref{sec:GenProp}. This work is summarized and its implications are presented in \S\ref{sec:CONC}.

\section{Observations and Data Reduction Procedures}
\label{sec:OBS&DATA}
\subsection{Selection of the Galaxy Sample}
\label{sec:OBS}
Our sample of galaxies was selected from the LSB galaxy catalog of Bothun et al.~(1985). Their work, based on the Uppsala General Catalog (UGC, Nilson 1973), classifies a galaxy as LSB if its average surface brightness $\mu_{\rm pg} \equiv m_{\rm pg} + 5\log(D) + 8.63$ (where $D$ is the galaxy diameter in arcmin and $m_{\rm pg}$ is the photographic magnitude, both from the UGC) satisfies $\mu_{\rm pg} \ge 25$ mag/arcsec$^{2}$ (Bothun et al.~1985; O'Neil et al.~2004). This definition is typically equivalent to the more commonly used classification of a galaxy as an LSB if the central surface brightness of the disk in the B-band, $\mu_{0,\rm disk}$, satisfies $\mu_{0,\rm disk} \ge 23$ mag/arcsec$^{2}$. The uncertainties in going from the equation which describes a mean surface brightness to $\mu_{0,\rm disk}$ are large though, and many galaxies satisfying the above criteria on $\mu_{\rm pg}$ can have $\mu_{0,\rm disk} \approx 22$ mag/arcsec$^{2}$ (O'Neil et al.~2004). We will come back to this issue in \S{3} below in order to clarify the general properties of our sample.

We restricted this study to galaxies from Bothun et al.~(1985) selected by their rotational velocity (measured through the 21-cm velocity width W20) to have baryonic masses comparable to that of the Milky Way ($M_{MW} \sim 7 \times 10^{11} M_\odot$). As shown below, this implies that these systems should have GC systems rich enough to be detected in our targets and provide information on their ensemble characteristics to contrast with the wealth of information on the properties of GC systems in HSB galaxies. The sample was further selected to span a range of $\sim 2$ magnitudes in surface brightness and we applied a redshift cut-off ($cz < 3000 $ km sec$^{-1}$ or $(m-M) \lesssim 33$ mag) to ensure that a large fraction of the GCs would be resolved using the ACS instrument on-board HST. Among the galaxies that satisfied these criteria, we selected 6 galaxies as a minimum number spanning the typical ranges of surface brightness, \ion{H}{1} mass, and $M_{\rm HI}/L_{B}$ ratios found in LSB galaxies.

It is well known that the number of GCs in HSB galaxies scales with the luminosity of the galaxy (e.g., Harris 2001). This scaling is usually quantified through the specific frequency $S_N$ (Harris \& van den Bergh 1981), which measures the number of GCs per unit $V$-band luminosity and is defined as $S_{N}=N_{GC} \ 10^{0.4(M_{V}+15)}$, where $M_{V}$ is the total visual magnitude of the galaxy. The specific frequency is observed to depend on Hubble type, with early-type systems having $S_N\sim 4$ and late-type systems $S_N\sim 1$, although the scatter around these values is significant. As no census of GCs in LSB galaxies exists, we conservatively assumed a low specific frequency of $S_N = 1$ in predicting the expected number of GCs in LSB galaxies. With this assumption, we expected between 50 and 150 GCs in our target galaxies.

A list of our target galaxies and  their basic properties is presented in Table~\ref{tab:basicdata} and a short discussion on the main features of each galaxy is included as an appendix. The distance to the galaxies was derived from the recession velocity corrected by Virgo infall using the local velocity field model given in Mould et al.~(2000) with the Hubble constant fixed to \mbox{$H_{0}=73\pm5$ km/s/Mpc}, and were obtained from the Nasa Extragalactic Database (NED). Absolute magnitudes (M$_{B_o}$) were derived with the most recent data compiled in NED and using the Schlegel et al.~(1998) reddening maps.

\begin{deluxetable*}{lrrrrrrr}
\tabletypesize{\scriptsize}
\tablecolumns{8}
\tablewidth{0pc}
\tablecaption{Properties of the galaxies.$^{a}$}
\tablehead{
\colhead{} & \colhead{UGC00477} & \colhead{UGC03459} & \colhead{UGC03587} & \colhead{UGC06138} & 
\colhead{UGC11131} & \colhead{UGC11651}& \colhead{MW}}
\startdata
$\alpha(J2000)$& 00:46:13 &   06:23:59 &  06:53:54 &   11:04:39 &   18:10:22 &   20:57:15 &      \\     
$\delta(J2000)$&+19:29:22 &  +04:42:39 & +19:17:59 &  +27:43:26 &  +01:35:33 &  +25:58:13 &      \\	
$l$    	      &    121.23 &     205.59 &    195.87 &     205.28 &      29.64 &      71.13 &      \\	
$b$	      &	   -43.36 &      -3.92 &      9.19 &      66.34 &       9.85 &     -12.58 &      \\     
$z$	      &  0.008836 &   0.009587 &  0.004226 &   0.008586 &   0.006014 &   0.005087 &      \\     
$v_r$ [km/s]  &      2698 &       2753 &      1227 &       2688 &       1958 &       1730 &      \\  	
$(m-M)$       &     32.84 &      32.88 &     31.13 &      32.83 &      32.14 &      31.87 &      \\   	
$d$ [Mpc]     &     36.98 &      37.67 &     16.83 &      36.81 &      26.79 &      23.66 &      \\	
$M_B$         &    -19.14 &            &    -19.10 &     -18.33 &     -19.58 &     -19.06 & -20.5\\ 	
$A_{V}$       &     0.120 &      1.860 &     0.306 &      0.115 &      1.610 &      0.879 &      \\	
E(B-V)        &     0.036 &      0.561 &     0.092 &      0.035 &      0.486 &      0.265 &      \\
$\mu_{m_{pg}}$[mag/arcsec$^2$]&     26.30 &    25.80 &     25.50 &      25.90 &      24.00 & 26.50 &  \\
$\mu_{0,\rm disc}$[mag/arcsec$^2$]& 21.86 &    21.43 &     22.63 &      23.74 &      21.62 & 21.75 &  \\
W20 [km/s]    &       246 &         311 &      220 &       224 &        206 &        276 &      \\
i [$^{\circ}$]&        13 &           53 &       22 &        31 &         26 &         15 &      \\
$D_{25}$ [\arcmin]  &    3.5 &       1.2 &        3 &         2 &        1.4 &        3.5 &      \\
$\log(M_{\rm HI}/M_{\odot})$ &  9.85 &   9.05 &   9.30 &       9.43 &       8.95 &        9.24& $\sim$9.7 \\
$\log(M_{\rm dyn}/M_{\odot})$ & 12.45 & 12.05 &  11.74 &      12.18 &      11.71 &	    11.97 & 11.77 \\	
Morphological Type   &	      Sdm &       Scd  &        S? &	     Sm &        Scd &        Sdm & Sb-Sc \\	
\enddata
\tablenotetext{a}{Positions, inclinations (i), 21cm velocity width (W20) and optical diameter at the 25mag/acrsec$^2$ isophote (D$_{25}$) were extracted from Karen O'Neil's Catalog of Massive LSB Galaxies (\textit{http://www.gb.nrao.edu/$\sim$koneil/biglsbgs/}) 
and were used to derive the masses (M$_{HI}$, M$_{dyn}$). The photographic surface brightness $\mu_{m_{pg}}$ was computed from Uppsala General Catalogue (Nilson, 1973). The central disk surface brightness in the B band $\mu_{0,\rm disc}$ was derived using the data presented in this work as described in the text. All the other parameters came either from NED or from HyperLeda databases. Distances and distance moduli are corrected by Virgo infall.}
\label{tab:basicdata}
\end{deluxetable*}

\subsection{Observations}

  Each galaxy was observed for a single HST orbit as part of program GO-10550 (PI: M.~Kissler-Patig). The Wide Field Channel of the Advanced Camera for Surveys (ACS; Ford et al.~1998) was used to obtain five $\sim$ 400s exposures: three in the F475W band ($\approx$ Sloan $g$) and two in the F775W band ($\approx$ Sloan $i$), which result in total exposure times of $\sim$1200 sec and $\sim$800 sec in the F475W and F775W bands respectively. In order to remove chip defects and bad pixels a line dither with a spacing of $0\farcs 146$ in the $g$-band and $0\farcs17$ in the $i$-band was performed in the identical exposures of each filter. A log of the observations is given in Table~\ref{tab:obslog}.

The choice of filters was dictated by the long baseline, which results in good sensitivity to metallicity and age of stellar populations, and by having at least one filter in common with the ACS Virgo and Fornax Cluster Surveys (C\^ot\'e et al.~2004; Jord\'an et al.~2007a). These surveys observed 143 early-type galaxies and their GC systems in the Virgo and Fornax clusters using F475W ($\approx$ Sloan $g$) and F850LP ($\approx$ Sloan $z$) and therefore matching one of their filters can prove useful to contrast the properties of our GCs with those of early-type galaxies. We chose F775W instead of F850LP due to its higher throughput. Note that here and throughout, we use $g$ as shorthand to refer to the F475W filter, and $i$ denotes F775W.

\begin{deluxetable}{ccc}
\tablecolumns{3}
\tablewidth{0pc}
\tablecaption{Observing log for GO-10550}
\tablehead{
\colhead{Galaxy} & \colhead{Exposure Time} & \colhead{Exposure Time}\\
\colhead{}       & \colhead{F475W}         & \colhead{F775W}}
\startdata
UGC~00477 &  3 $\times$ 404  &  2 $\times$ 405  \\
UGC~03459 &  3 $\times$ 400  &  2 $\times$ 400  \\
UGC~03587 &  3 $\times$ 404  &  2 $\times$ 405  \\
UGC~06138 &  3 $\times$ 406  &  2 $\times$ 407  \\
UGC~11131 &  3 $\times$ 400  &  2 $\times$ 400  \\
UGC~11651 &  3 $\times$ 406  &  2 $\times$ 407  \\
\enddata
\label{tab:obslog}
\end{deluxetable}

\subsection{Data Reduction and Object Detection}
\begin{figure*}
\begin{center}
\includegraphics[width=15cm]{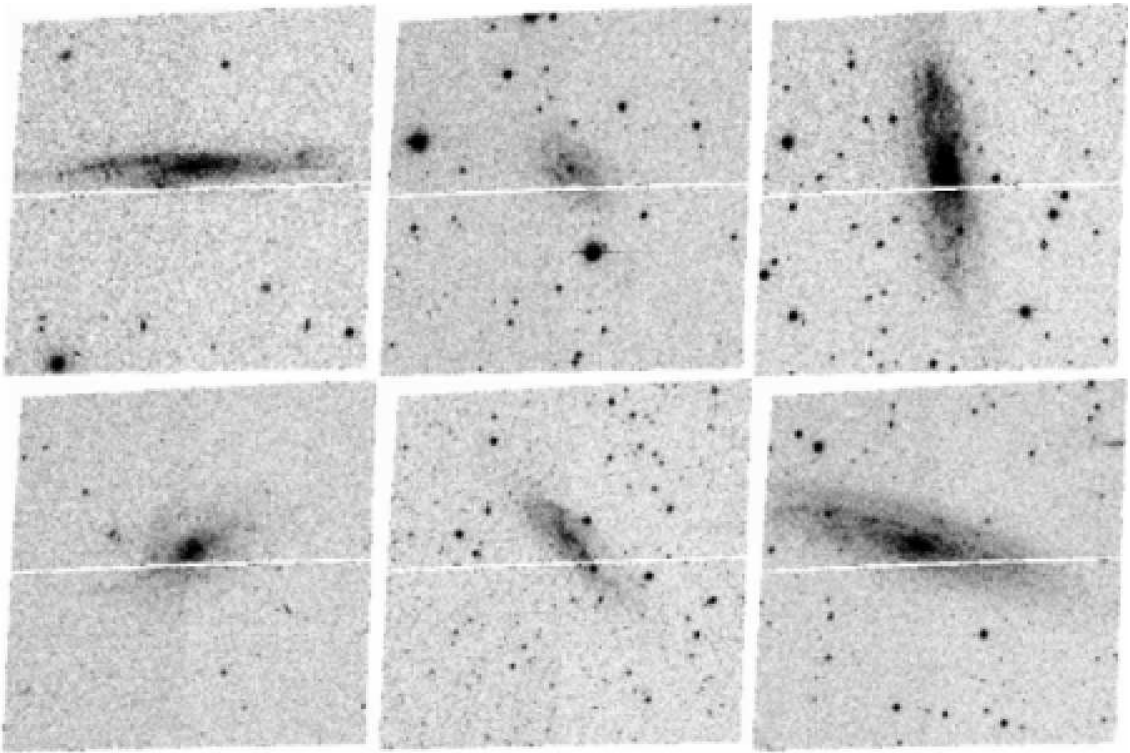}
\caption{F475W band images of the six observed galaxies. From left to right, upper panel:  \textit {ugc00477}, \textit{ugc03459}, \textit{ugc03587}; lower panel: \textit{ugc06138}, \textit{ugc11131}, \textit{ugc11651}.}
\label{fig:images}
\end{center}
\end{figure*}

The images were combined and cosmic-ray cleaned using the task ``multidrizzle'' in Pyraf (Koekemoer et al.~2002) without performing sky subtraction and by adopting a ``lanczos3'' kernel function for the final image combination. In order to register the images the header information was used, after checking using foreground stars in our frames that the actual shifts between the images were consistent with the commanded ones. The output of the image combination is an exposure time weighted image of $4218\times 4243$ pixel$^2$ in each filter for each galaxy. Figure~\ref{fig:images} shows the co-added F475W image for the six galaxies studied in this work.

Object detection was carried out with SExtractor (Bertin \& Arnouts 1996). In order to aid in the object detection and background determination a weight image was created using the WHT extension of the ``multidrizzle'' final image, following the procedure outlined by Jord\'an et al.~(2004). The result of this process, a map of the {\it rms} in the image, was used when running SExtractor by setting the ``WEIGHT\_TYPE'' parameter to ``MAP\_RMS''. Object detection was performed in each image independently requiring a minimum of 5 connected pixels above a threshold of 2$\sigma$. The background was estimated using a mesh size of $40\times 40$ pixel$^2$ with no background filtering. The resulting object catalogs in each filter were matched using a matching radius of $0\farcs 1$ and only sources that were matched were retained for further analysis.

Aperture photometry was performed using an aperture radius of 4 pixels, with the background estimated locally in a rectangular annulus with a width of 30 pixels. The apparent AB magnitudes of the objects were calculated from the instrumental magnitude using the photometric zeropoints and aperture corrections given by Sirianni et al.~(2005)%
\footnote{The distances to our target galaxies imply that many of the GCs should be marginally resolved given their typical half-light radii $r_h \sim 3$ pc (e.g., Jord\'an et al.~2005). This implies that the aperture corrections for point sources will in general underestimate the correction needed for GCs. We have checked the magnitude of the expected difference by creating simulated GCs as point-spread function convolved King (1966) models with half-light radius $r_h=3$ pc and concentration $c=1.5$, i.e.~a model appropriate for a typical GC. We find that the average expected corrections vary by $\lesssim 0.02$ mag in $g$ and $\lesssim 0.05$ mag in $i$. Given the modest size of the expected corrections for an average object we have chosen to adopt the corrections listed in Sirianni et al.~(2005) for all target galaxies. Although such offsets do not affect our conclusions in any significant way, it should be kept in mind that magnitudes and colors can have systematic offsets $\lesssim 0.05$ mag due to the fact that GCs are marginally resolved.}. 
The dust maps of Schlegel et al.~(1998) and the extinction ratios listed in Sirianni et al.~(2005) were used to correct the observed magnitudes by foreground Galactic absorption. No corrections were applied to account for possible reddening internal to the target galaxies.

Finally, we fit point-spread function (PSF) convolved King (1966) models to all detected sources using the code described in Jord\'an et al.~(2005). The code returns the best-fit concentration $c$ and half-light radius $r_h$ for each source, but we use only the latter as the former is poorly constrained. The half-light radii are useful in selecting bona-fide GC candidates from the full list of detections (see below).

\section{Surface Brightness Profiles and Galaxy Classification}
\label{sec:SBP}
In order to obtain an improved classification of the target galaxies based on the central surface brightness we used our observations to derive surface brightness profiles. We then used double S\'ersic model fits to describe the profiles and classify them according to their derived values of $\mu_{0,\rm disk}$, the central surface brightness of the large scale (disk) component of the profile. In this section we describe these measurements and our new classifications.

\subsection{Surface Photometry}
An isophotal analysis of the light profile of the target galaxies was performed running the IRAF task {\it ``ellipse''} which is based on the algorithm of Jedrzejewski (1987). This task was run on a version of our ACS images where bright sources such as foreground stars were masked. We used fixed position angles and ellipticities, which were estimated after some experimentation with {\it ``ellipse''}, and allowed 20 pixel as the maximum wander between successive isophote centers. Using this information we obtained an azimuthally averaged intensity profile for each galaxy. The ``sky" brightness was determined by measuring the emission level in five boxes of $200\times200$ pixel$^2$ located in regions of the image that appear to be free of galaxy light. The average of the mean counts in the five boxes was adopted as the background value for every image and the standard deviation between these values was used to estimate the errors in the sky determination.

\begin{figure*}
\begin{center}
\includegraphics[angle=0,scale=.7]{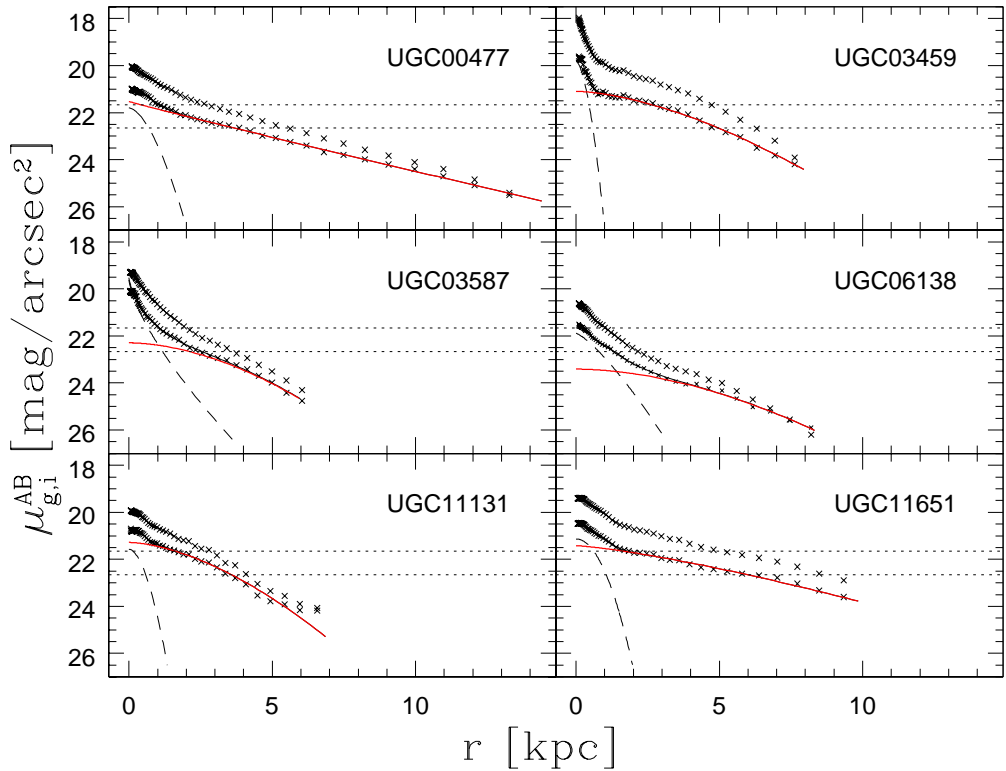}
\caption{Surface brightness profiles in the $g$- (down) and $i$-bands (up) for the target galaxies. The red-solid line shows the best S\'ersic profile fit to the large (disk) component and the dashed line shows the  corresponding inner (bulge) S\'ersic component, both for the $g$-band profile. The two horizontal lines indicate the limits for normal to intermediate and intermediate to low surface brightness and are located at $\mu_{B}$=22 mag/arcsec$^2$ and  $\mu_{B}$=23 mag/arcsec$^2$ and translated into the $g$-band by assuming $(g-B)=-0.34$ (Fukugita et al.~1995).} 
\label{fig:lp}
\end{center}
\end{figure*}

The measured intensity profiles were converted to AB magnitudes per square arcsecond using the photometric zeropoints of Sirianni et al.~(2005) and a pixel size of $0\farcs05$. The profiles were corrected for foreground extinction by using $E(B-V)$ values from Schlegel et al.~(1998) and the extinction ratios listed in Sirianni et al.~(2005). The resulting surface brightness profiles in the $g$- and $i$-bands are shown in Figure~\ref{fig:lp}.

\subsection{Model Fits and Galaxy Classification}

In order to estimate the surface brightness of the disk $\mu_{0,\rm disk}$ of our target galaxies we decomposed them by fitting two S\'ersic (1968; see Graham \& Driver 2005 for a review) profiles, one to describe a bulge component and the other to describe a larger scale disk component. A S\'ersic profile is given by

\begin{equation}
I(r) = I_e \exp[-b_n((r/r_e)^{1/n}-1)]
\end{equation}

\noindent where $n$ is the S\'ersic index, $I_e$ the intensity of the profile at the effective radius $r_e$ and $b_n$ ensures that $r_e$ contains half of the integrated light and can be approximated by $b_n \approx 1.992n-0.3271$ (see Graham \& Driver 2005 and references therein).

It is usual to model disks as pure exponential profiles (or S\'ersic profiles with $n\equiv 1$) but we found that our profiles are usually not well described by exponential profiles on large scales. This is consistent with the poor description afforded by exponential fits to describe the large scales of a sample of 21 late-type galaxies with a range of surface brightness (Galaz et al.~2006; their online Figures 2 and 3). We have therefore chosen to use the more general S\'ersic profile to describe the disk component as this can properly describe the data. In all cases the best-fit S\'ersic component describing the disk satisfies $n \lesssim 1$. We carried out the fits applying a $\chi^2$ minimization using procedures from CERN's Minuit library. The minimization procedure used consists of a Simplex minimization algorithm (Nelder \& Mead 1965) followed by a variable metric method with inexact line search (MIGRAD) to refine the minima found by the Simplex algorithm. Following Byun et al.~(1996) and Ferrarese et al.~(2006; their equation 14) we weighted all points equally in a fractional sense. Our best fit double S\'ersic models are shown in Figure \ref{fig:lp} and they reproduce the data very closely. The external S\'ersic component is taken to be the disk and to infer the value of $\mu_{0,\rm disk}$ in the $g$-band. In order to estimate $\mu_{0,\rm disk}$ in the $B$-band, we assumed a typical color $(g-B) = -0.34$ mag for an Sbc galaxy (Fukugita et al.~1995). The values we inferred for the B-band value of $\mu_{0,\rm disk}$ are then 21.86 (\object[UGC 00477]{{\it ugc00477}}), 21.43 (\object[UGC 03459]{{\it ugc03459}}), 22.63 (\object[UGC 03587]{{\it ugc03587}}), 23.74 (\object[UGC 06138]{{\it ugc06138}}), 21.62 (\object[UGC 11131]{{\it ugc11131}}) and 21.75 (\object[UGC 11651]{{\it ugc11651}}). We used these values to re-classify our target galaxies as described below.

There is no clear definition of the limiting surface brightness separating low from high surface brightness galaxies. The convention has been evolving through the years as much as the limiting magnitude of the available galaxy surveys did. The most recent works on low surface brightness galaxies are mostly restricted to galaxies with central surface brightness of $\mu_{B}\gtrsim$ 23 mag/arcsec$^{2}$ (e.g., Zackrisson et al.~2005), but several different definitions are also used, one of the most common ones being a blue central disk surface brightness value of $\mu_{0,\rm disk}\gtrsim$ 22 mag/arcsec$^{2}$ (e.g.~Boissier et al.~2003). As shown in Figure \ref{fig:lp} the observed values of $\mu_{0,\rm disk}$ imply that our sample covers a large range of central surface brightness. We have used the model fits described above to re-classify our target galaxies following McGaugh et al.~(1995), which defines LSBs as those objects showing a central disk surface brightness in the B band of $\mu_{0,\rm disc} >$ 23 mag/arcsec$^{2}$ and ``intermediate surface brightness'' those galaxies satisfying 22 mag/arcsec$^{2}$ $<\mu_{0,\rm disc}<$ 23 mag/arcsec$^{2}$. Using these definitions we classify \object[UGC 06138]{{\it ugc06138}} as an LSB and \object[UGC 03587]{{\it ugc03587}} as an intermediate surface brightness galaxy. The other 4 galaxies in our sample are ``normal'' (or high) surface brightness galaxies, despite having low mean surface brightness. 

The sample of Bothun et al.~(1985) from which our sample was drawn is selected on mean surface brightness, which usually translates into $\mu_{0,\rm disc} \gtrsim 23$ mag/arcsec$^{2}$ but with objects with values of $\mu_{0,\rm disc}$ as high as 22 mag/arcsec$^{2}$ (O'Neil et al.~2004). Our findings are roughly consistent with this view when taking into account that the four galaxies that fall into the high surface brightness category suffer from a high estimated extinction in their field of view. 

\section{Globular Cluster Candidate Selection Criteria}
\label{sec:GCsel}

  The SExtractor output catalogs of objects in each field are strongly contaminated by both foreground Galactic stars and background galaxies. Therefore, we need to define criteria in order to identify potential globular clusters from the full list of detections. We selected our GC candidates as those objects that fulfill the following conditions:

  \emph{Magnitudes}. The GC luminosity function (GCLF; the distribution of GCs per unit magnitude) is observed to have a nearly gaussian shape with a roughly constant peak at a magnitude of $M_V\sim -7.4$ mag (Harris 2001), which corresponds to $M_g \sim -7.1$ mag according to stellar population models for a single-burst population of 10 Gyr with [Fe/H]$=-1.35$ (Maraston 2005). For a galaxy having a mass like the Milky Way as our targets, the typical dispersion $\sigma$ in a Gaussian description of the GCLF is $\sigma \sim 1.2$ mag (Jord\'an et al.~2006; 2007b). Therefore, we adopted an upper luminosity cut at $2.5\sigma$ above the expected GCLF turnover in the $g$-band, which translates into the condition $M_g > -10$ mag. 

At low GC luminosities we impose a cut on apparent magnitude driven by the completeness limit of the observations. The completeness function depend both in the magnitude of the objects and their position on the galaxy. To account for these effects we added to the galaxy fields simulated GCs with a given magnitude and a typical size of 3~pc, randomly distributed along the image. The photometry was then performed in the same fashion as for the real observations. After that the number of fake added sources recovered was computed. The iteration of this process in steps of 0.01 magnitudes at least 50 times per field of view allowed us to determine the 100\% completeness limit of our observation with 10\% confidence level. These limits correspond to 26.3 mag and 25.5 mag in the $g$- and $i$-bands respectively and were used to fix the low magnitudes cut of our sample selection. 

  \emph{Color}. A broad color range, $0.4<(g-i)_{0}<2.0$, was used in order to include GCs with metallicities satisfying $-2.26 < $ [Fe/H] $ < 0.35$ for an assumed old $\tau=10$ Gyr age and to allow some intrinsic reddening in the target galaxies. This color cut corresponds to $0.42 \lesssim (B-V)\lesssim 1.43$, a range that includes 84\% of the galactic GCs with available $(B-V)_{o}$ color information in the McMaster catalog of Milky Way GCs (Harris 1996). 

  \emph{Morphology}. Observations in the Milky Way show that GCs are fairly spherical systems (White \& Shawl 1987). The most extremely flattened Galactic case, NGC~6273, has an ellipticity of 0.27, which corresponds to an axial ratio of 1.33. Based on these observations we included in our catalog only objects whose SExtractor's measured axial ratio was $a/b<1.5$, leaving an wide range for more flattened objects but discarding mostly background galaxies. We note that due to the fact that GCs in our target are marginally resolved, the adopted condition is very conservative as the observed axial ratios are strongly affected by the PSF and therefore high axial ratios will tend to be circularized. 

We also included a second morphological selection criteria based on SExtractor's ``CLASS\_STAR'' parameter, which on the basis of a trained neural network assigns to each source a value between 0 (galaxy) and 1 (star) (see Bertin \& Arnouts 1996 for details). Objects with CLASS\_STAR $< 0.1$ were excluded from the final catalog.

  \emph{Size}.  Using the measured half-light radii $r_h$ we exclude from the final catalog all objects that have a half light radii in the $i$ band $r_{h, i}<0.25$ pix. This rejects sources that are consistent with being point sources, i.e.~mostly stars, and therefore eliminates most of the foreground contamination in our sample. For the most distant galaxies ($D\sim 35$ kpc) this cut corresponds to $r_{h,i} \sim 2.2$ pc which would exclude $\sim 25\%$ of the total globular clusters population of the Milky Way if observed at those distances because of being more compact than our limit. In the case of the closest galaxy ($D= 16.83$ kpc) this value goes down to 1 pc which would exclude just a few ($\sim$ 5) Galactic GCs. 

We adopted a physical upper limit to the radius of $r_{h, i}<10pc$, since just $\sim$5\% of the Milky Way globular cluster systems have $r_{h, i}$ higher than this value. The upper cut in $r_h$ implies that our study does not probe for the existence in our sample galaxies of objects such as diffuse star clusters or faint fuzzies (Brodie \& Larsen 2003; Peng et al.~2006) and ultra-compact dwarfs or dwarf-globular transition objects (Hilker et al.~1999; Drinkwater et al.~2000; Ha\c{s}egan et al.~2005).

As a final requirement on the structural parameters, we eliminated objects whose fractional uncertainties $\delta r_{h} \equiv \sigma_{r_{h}} / r_{h}$, where $\sigma_{r_{h}}$ is the uncertainty in the measured $r_h$, satisfy $\delta r_{h} \ge 0.3$ in any of the bands. A large error in the measured $r_h$ is an indication that the observed light distribution is not properly fit by a King (1966) model, and as a consequence it is likely that the source is a background galaxy instead.

\begin{figure*}
\begin{center}
\includegraphics[angle=0,scale=.8]{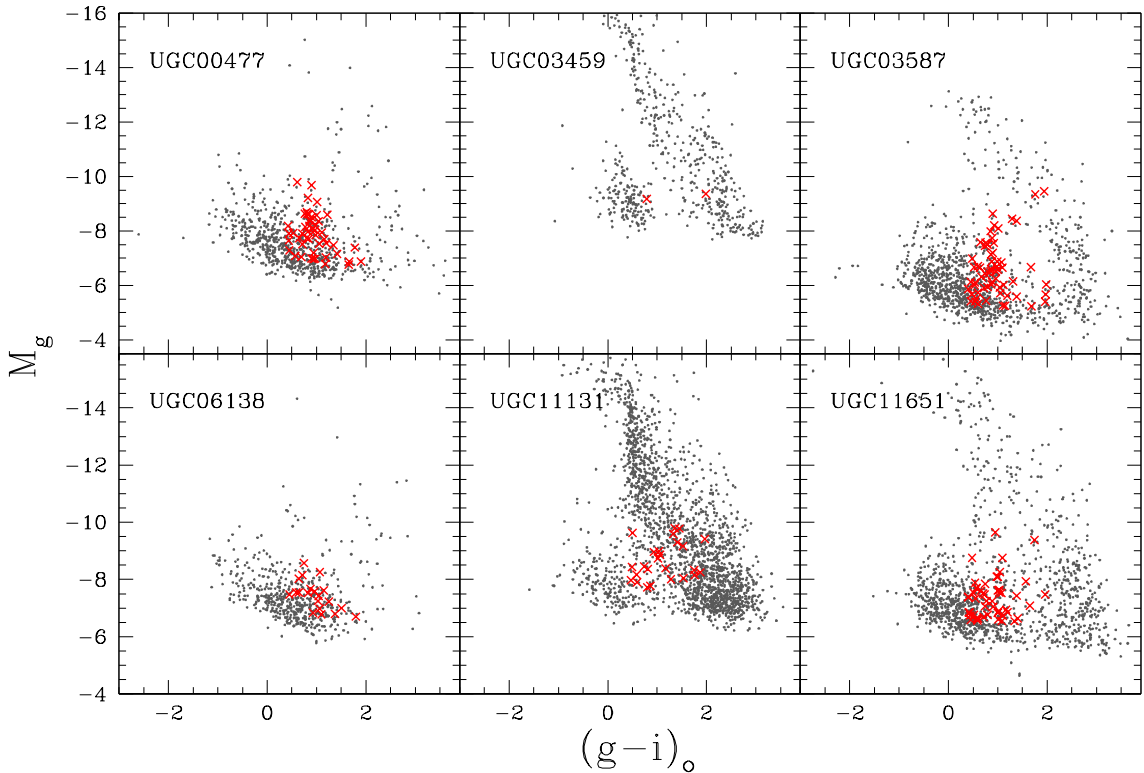}
\caption{Color-magnitude diagrams of all the sources detected in the field of view of each galaxy (small background dots). The red crosses correspond to those sources classified as globular cluster candidates. } 
\label{fig:cmd}
\end{center}
\end{figure*}

Figure~\ref{fig:cmd} shows the location in a color magnitude diagram of objects classified as globular cluster candidates according to the criteria described above (red crosses). The small background dots in the same diagram represent all additional sources detected in the field of view of each galaxy.

\subsection{Control Fields} 

Even though we expect to obtain a fairly clean catalog of GC candidates after applying the selection criteria described above, some contamination will remain in the sample. In order to estimate the expected level of remaining contamination in the object sample of each galaxy, we searched the HST/ACS archive for blank fields at least as deep as our observations, and whose galactic coordinates would make them good control samples, either because they are located close to one of our fields, or in a symmetric position with respect to the Galactic center. These conditions were chosen such that the properties of the galactic foreground in the control fields  were similar to our observed frames, but note that given that we can eliminate most stars using our half-light radii measurements a very accurate match is not crucial. 
Two of the galaxies,  \object[UGC 00477]{{\it ugc00477}} and  \object[ugc06138]{{\it ugc06138}}, are located at very high galactic latitude, therefore any high galactic latitude blank field is suitable to reproduce the contamination as in this case the amount of foreground Galactic stars is very low. For these galaxies we were able to obtain six control fields which are denoted C01 to C06 in Table~\ref{tab:control_fields}. The contamination for these two galaxies was estimated as the average expected number of contaminating objects in the six control fields.

\begin{deluxetable}{rrrcc}
\tablecolumns{6}
\tablewidth{0pc}
\tablecaption{Control Fields}
\tablehead{
\colhead{ID} & \colhead{\textit{l}} & \colhead{\textit{b}} & \colhead{Prog. ID} & \colhead{E(B-V)$^{a}$}}
\startdata
C01 &    145.58 & -38.65 & 9488  & 0.155 \\   
C02 &    115.01 &  46.68 & 9488  & 0.018 \\   
C03 &     99.41 & -47.15 & 9488  & 0.033 \\   
C04 &    289.08 &  62.25 & 9488  & 0.024 \\   
C05 &    251.33 & -41.44 & 9488  & 0.008 \\   
C06 &     92.66 &  46.37 & 9488  & 0.016 \\   
C07 &    279.93 & -19.99 & 9575  & 0.130 \\   
C08 &    179.06 & -19.93 & 9488  & 0.630 \\   
C09 &    105.10 &   7.07 & 9488  & 0.824 \\   
C10 &     24.48 &  12.67 & 9488  & 0.478 \\   
\enddata
\tablenotetext{a}{Foreground Galactic reddening in the direction of each field of view from Schlegel et al.~1998.}
\label{tab:control_fields}
\end{deluxetable}

The other four galaxies in the sample have low galactic latitude, making the selection of control fields harder as not many suitable observations have been carried out in such regions with the filter set used in this program. In these cases we had to restrict ourselves to finding a control field as close as possible to our target, which meant using just one or two control fields per galaxy. Column 2 in Table~\ref{tab:SN} indicates the ones we consider to be the best control field for a given galaxy according to the labels used in Table~\ref{tab:control_fields}.

All the chosen control fields were observed as part of two ACS Pure Parallel programs, GO-9488 and GO-9575, and they have exposure times $1600 \mbox{ sec} \leq t \leq 1800 \mbox{ sec}$ in both bands. These images were reduced and the photometry was performed using exactly the same procedures followed for our targets. In order to account for the influence of the light of each target galaxy on the detections over our frames we follow a procedure similar to that described in \S2.2 of Peng et al.~(2006a). Namely, for each galaxy we created ``custom'' control samples by re-doing the detections {\it as if} a given galaxy was in front of our blank control fields. To do this, we used the noise images created for each galaxy in the data reduction process  when running the detection procedure on the control fields. In this way we can reproduce for each control field the catalog of objects that we would have detected if there had been a particular galaxy of our sample in the field.

The same selection criteria described above were applied to the final SExtractor catalogs of the control fields in order to obtain a final catalog of the expected contaminating objects. As we aimed at reproducing the expected contamination in the case the control field was in the same position as our target fields, the apparent magnitude for the control objects was computed as:

\begin{equation}
       m= m_{obs}- A_{cf} + A_{gf}
\end{equation}

\noindent where $m_{obs}$ is the observed magnitude in a given band and $A_{cf}$ and $A_{gf}$ are the extinction corrections for the control field and the galaxy field in the same band respectively.

\section{Total Number of Globular Clusters}
\label{sec:TotalGC}
Once the selection criteria have been applied to all the objects in each field we have the total number of observed GC candidates. In order to have an accurate estimate of the total population of GCs in our target galaxies, we need to consider both the expected number of contaminants and the fraction of objects we might be losing due to observational and selection effects. Below we describe our adopted procedures to account for these observational effects.

\subsection{Contamination}

There are mainly two different kinds of contamination in our sample. One is {\it external} in nature and is composed mostly of background galaxies and any foreground star whose inaccurate $r_h$ measurement might have allowed it to pass our rejection of unresolved sources. This external contamination can be corrected by using the control fields customized to the field of each target galaxy. Another set of contaminants is of {\it internal} nature, and is composed of objects in each target galaxy which given our adopted selection criteria in morphology, color, magnitude and size are almost certainly reddened star forming complexes.

The typical size for an ultra compact \ion{H}{2} region is $r \lesssim 1.6$ pc (Churchwell 2002). Objects with these sizes might be present in our sample in the three closest galaxies, and this could constitute a source of contamination, as some objects with those sizes are indeed detected (see Figure~\ref{radio} and discussion below). The other main expected contribution from non-GC sources in the target galaxies to our candidate sample will be due to reddened young star clusters lying in the spiral arms of our targets. Young star clusters typically show sizes between 1 - 20 pc, with typical radii of $r \sim 3$ pc (e.g., Scheepmaker et al.~2007), magnitudes satisfying $  M_V \gtrsim -11.5 $ and colors in the range $-0.2 \lesssim (B-V) \lesssim 0.4$ (Larsen 1999), roughly corresponding to $-0.57 \lesssim (g-i) \lesssim 0.36$. Therefore, even a small amount of reddening may allow young star clusters to be included in our sample of (old) GC candidates. Visual inspection of our images shows that most of our GC candidates are {\it not} located in obvious star forming regions, even though a small amount of contamination is still possible. 

We show in \S\ref{sec:GenProp} that the properties (luminosity function, color and size distribution) of the combined sample of GC candidates in our sample of LSB galaxies are consistent with the expected properties of old GCs in HSB galaxies (see below). Therefore,  we are confident that our sample of GC candidates is not dominated by young star clusters. Of course, while the total sample is likely dominated by old GCs, the nature of any given object can only be assessed by spectroscopic follow-up.

\subsection{Incompleteness}

Due to observational effects and the various selection criteria we are using, we do not expect to be able to observe the complete GC system of each galaxy. Therefore, we need to estimate the fraction of the GC system we might be missing and use this information to estimate the total size of the GC systems.
As the Milky Way has the best studied GC system and its size and morphology are comparable to those of our target galaxies, we used the observed properties of its GC system in order to estimate the total number of clusters in each galaxy through a direct comparison.
In order to do so, we follow a procedure similar to the one described by Kissler-Patig et al.~(1999). Based on the areal coverage of the WFC we create a spatial ``mask'' for each galaxy and we apply this mask and our additional selection criteria to a projection of the Galactic GC system as described below. The properties of the Galactic GC systems are obtained form the McMaster catalog (Harris 1996).

We apply our selection criteria to the two-dimensional spatial distribution obtained projecting the Galactocentric coordinates $(X,Y,Z)$ of the Galactic GC system into the $Y$-$Z$ plane and rejecting sources that lie outside the spatial mask of each galaxy and that do not satisfy the selection criteria presented in \S\ref{sec:GCsel}. The total number of GCs for a given galaxy, $N_{GC}$, is then given by

\begin{equation}
       N_{GC}=N_{MW} \frac{N_{CC}}{N_{\rm mask}},
\end{equation}

\noindent where $N_{CC}$ is the number of GC candidates in each target galaxy left after subtracting from the observed number of GC candidates the external contamination, $N_{\rm mask}$ is the number of Galactic clusters detected inside the mask and $N_{MW}$ is the total number of clusters in the Milky Way which we assume to be 180$\pm$20 (Ashman \& Zepf 1998). In order to apply to the Milky Way GCs the same selection criteria we apply to our target galaxies, the cuts in magnitude were applied to
absolute magnitude of the Galactic clusters that were reddened as appropriate for each of our
target galaxies, and 
$(B-V)$ colors were transformed to $(g-i)$ using a linear conversion 
derived from the Maraston (2004) 
models\footnote{The transformation used was $(g-i) = 1.57(B-V) + 0.26$.}.

This direct comparison with the GC system of our Galaxy relies on several assumptions which we now discuss.

\subsubsection{The Galactic GC sample}

The McMaster Catalog lists a total of 150 globular clusters, all of them with Galactocentric coordinates. There are four GCs without photometric information, so they are excluded from the comparison sample and we start working with 146 object, accounting the 4 missing clusters as part of the uncertainty in the total number of GCs in the Galaxy. From these 146 GCs, 29 have no color information and, from the remaining sample, 17 do not have measured ellipticities. Therefore, the comparison sample is limited to 100 Galactic GCs. 

The exclusion of these 46 clusters does not represent a significant bias in the case of  \object[UGC 00477]{{\it ugc00477}}, \object[UGC 03459]{{\it ugc03459}}, \object[UGC 06138]{{\it ugc06138}} and \object[UGC 11131]{{\it ugc11131}}, because at the distances and reddening factors of these galaxies those objects would be rejected anyway either by the faint magnitude cut or by the color/ellipticity selection, i.e.~they are rejected without the need to use the missing information. For \object[UGC 03587]{{\it ugc03587}} and \object[UGC 11615]{{\it ugc11651}} we face a different situation, because after applying all the selection criteria, except the one that uses the missing quantity to these 46 objects, 14 and 5 objects are still left for those two galaxies respectively. These numbers constitute an additional source of uncertainty for these galaxies.

Following Ashman \& Zepf (1998), we assume that the Milky Way has a total of 180$\pm$20 GCs, where the unobserved clusters are assumed to be hidden behind the Galactic bulge. We assume in what follows that in our target galaxies the fraction of obscured GCs is the same on all galaxies and is given by the Galactic estimate of this fraction.

\subsubsection{The Magnitude Distribution of the Milky Way GCs}

When setting up the selection criteria and when comparing the GC population of the target galaxies with the Milky Way's GC system we are implicitly assuming that the luminosity function of both systems is similar. 
It has been recently shown that the GCLF depends systematically on the galaxy mass (Jord\'an et al.~2006; 2007b), but given that all of our target galaxies have similar masses between them and with the MW we can safely assume that the GCLFs of our target galaxies are roughly similar than that of the MW. 
In order to make a consistent comparison between the systems, we need to make sure to consider the same fraction of the GCLF in the Milky Way and any sample galaxy. In order to do that the cut at faint magnitudes was applied over the (un-de-reddened) absolute magnitudes of the Galactic clusters and then dimmed by the distance and absorption coefficients corresponding to each galaxy. 
At bright magnitudes the cut was applied at M$_V\sim -10.3$. 

\subsubsection{The Spatial Distribution of the Milky Way GCs}
 
Another implicit assumption done when using the spatial mask is that the spatial distribution of the Milky Way GC system is similar to that of the GC systems of our target galaxies. 
Little is known about the spatial distribution of GCs in spiral galaxies. In general, the spatial distributions seem to be properly fitted by a projected power law with indices ranging from -2 to -2.5, but some differences exist between Milky Way-like galaxies with a more concentrated distribution, and M31-like galaxies where the GC system is more extended than the halo light distribution (see Ashman \& Zepf 1998 for a review). Harris (1986) and Kaisler et al.~(1996) have shown that more luminous galaxies have shallower radial distributions.

If the sample galaxies had much steeper profiles as compared to the one observed in the Galaxy we would be overestimating the number of clusters that are left out of the mask, but this would not greatly affect the final measurement, because even for the galaxy with the smaller areal coverage (\object[UGC 03587]{{\it ugc03587}}) the mask includes more than half of the the Galactic globular clusters, when assuming it has a similar profile. If the spatial distribution of GCs in the galaxies happen to be shallower than the observed in the Milky Way we will be accordingly underestimating the total number of clusters of the system.

\subsubsection{Distances}
 
The uncertainty in the distances to the galaxies is likely to be the most important systematic error in this study, because distances are involved in the absolute magnitude and size selection and they are also crucial to accurately reproduce the field of view for each galaxy when creating the comparison mask. The distances we used are corrected for the Virgo infall, but still have the intrinsic error due to the peculiar velocity of field galaxies and the uncertainty in the value of $H_0$, which we assume to be 73 km/s/Mpc. Estimates of the peculiar velocity for field galaxies depend on the method and the sample used, but they are thought not to exceed $\sim 100$ km sec$^{-1}$ (see, e.g., Davis et al.~1997). 

As no better distance estimate of our target galaxies is available, we evaluated how our results change under a $\pm$15\% variation in the assumed distances. This exercise is then used to account for the uncertainty due to peculiar velocities and the error in $H_0$.  

\subsection{The Total Number of GCs}

We determined the total number of clusters for each galaxy following the method described above. Table~\ref{tab:SN} shows our results and their uncertainties using our assumed distances to the galaxies as well as the results with these distances varied by $\pm$15\%. The uncertainties reported in Table~\ref{tab:SN} include the random errors described above estimated assuming Poisson statistics for the number of GC candidates, the expected contamination in each field and the number of objects inside the Milky Way's mask.

Additionally, several sources of systematic uncertainties need to be accounted for. To the uncertainty in the distances (which introduces significant differences in the final number of clusters), we need to add the 14 and 5 Galactic GCs lacking information in the case of \object[UGC 03587]{{\it ugc03587}} and \object[UGC 11651]{{\it ugc11651}}, respectively. The difference in the final GC numbers due to these missing objects should not exceed -22/+32 and -6/+6 objects for each of these galaxies, respectively. Furthermore, we assumed a fraction of $\sim20\%$ of obscured clusters in both galaxies even though this may not be the case. Finally, but certainly not less important, four galaxies in the sample are located at very low galactic latitudes, where we have to work with large and uncertain extinction values. Especially critical is the case of \object[UGC 03459]{{\it ugc03459}}, where the high galactic extinction combined with the distance to the galaxy led in practice to a result consistent with a null detection of cluster candidates.

\section{General Properties of the Globular Cluster Systems}
\label{sec:GenProp}
In what follows we present general properties of the GC systems detected in our program galaxies. Whenever appropriate, we divide the sample according to our derived values of $\mu_{0, \rm disk}$ in order to probe for any effect that surface brightness might have on the properties under study. In all the following sections except in \S\ref{sec:SpecFreq}, we refer to the properties of all the 206 objects classified as globular cluster candidates, without performing a statistical subtraction of the expected contamination. 
   
\subsection{Specific Frequencies}
\label{sec:SpecFreq}

We present in Table~\ref{tab:SN} the derived specific frequencies for our target galaxies, which are defined as $S_{N}=N_{GC} \ 10^{0.4(M_{V}+15)}$ (Harris \& van den Bergh 1981), where $M_{V}$ is the total visual magnitude of the galaxy which was transformed from the known $M_{B}$ magnitude assuming a typical color for an LSB galaxy of $(B-V)=0.6$ mag (see Table~2 in de Blok et al.~1995). The statistical error was computed in this case assuming a 0.2 mag uncertainty in the magnitude of the galaxy. 

Within the uncertainties the results are consistent with the typical $S_{N}$ for late-type galaxies, i.e.~$S_N \sim 1$. The average $S_N$ for the 5 sample galaxies with GC detections is $S_N = 2.34 \pm 0.99$, which is slightly higher than the average $S_N$ for late-type galaxies. McLaughlin (1999) suggested from studies in early-type galaxies that there is a universal GC formation efficiency by mass $\epsilon_{\rm cl}$ (i.e.~the fraction of the total baryonic mass that turns into GCs; see Blakeslee 1999 for a similar proposal), and inferred a value of $\epsilon_{\rm cl} \approx 0.25\%$. The fact that the specific frequencies we measured are a factor of $\sim 2$ higher than those typical in normal brightness late-type galaxies is consistent with the idea of a universal GC formation efficiency given the fact that galaxies classically classified as having low surface brightness have mass-to-light ratios that are $\approx 2$ times higher than of higher surface brightness galaxies (Zwaan et al.~1995; Impey \& Bothun 1997). In particular, the specific frequency for our two lowest surface brightness galaxies (\object[UGC 03587]{{\it ugc03587}} and \object[UGC 06138]{{\it ugc06138}}) is $\sim 2$, albeit with large uncertainties.

\begin{deluxetable*}{ccccrccrccrc}
\tabletypesize{\scriptsize}
\tablewidth{0pc}
\tablecaption{Number of globular clusters and upper limit for the specific frequency per galaxy.$^{a}$}
\tablehead{
\colhead{} & \colhead{} & \colhead{} & \colhead{} & \colhead{}& \colhead{} & \colhead{} 
&\multicolumn{2}{c}{15\% closer} & \colhead{} & \multicolumn{2}{c}{15\% farther} \\
\cline{8-9}  \cline{11-12}\\
\colhead{Galaxy} & \colhead{Control Field(s)} & \colhead{$N_{CC}$} & \colhead{$N_{MW}$} & 
\colhead{$N_{GC}$} & \colhead{$S_{N}$}& \colhead{}& \colhead{$N_{GC}$} & \colhead{$S_{N}$}& \colhead{} 
& \colhead{$N_{GC}$} & \colhead{$S_{N}$}\\
\colhead{(1)} & \colhead{(2)} & \colhead{(3)} & \colhead{(4)} & \colhead{(5)} & \colhead{(6)}& \colhead{}
& \colhead{(7)} & \colhead{(8)}& \colhead{} & \colhead{(9)} & \colhead{(10)} }
\startdata
UGC~00477& C01-C06 &42 (49)& 31 & 244 $\pm$ 80 & 3.10 $\pm$ 1.17 & & 172 $\pm$ 54 & 2.27 $\pm$ 0.83 & & 263 $\pm$ 94 & 3.46 $\pm$ 1.39 \\
UGC~03459& C08,C09 &\ 0 \ (2)&  6 & --- $\pm$ ---&  --- $\pm$ ---  & & --- $\pm$ ---&  --- $\pm$ ---  & & --- $\pm$ ---& --- $\pm$ --- \\
UGC~03587& C08,C09 &57 (61)& 57 & 180 $\pm$ 43 & 2.37 $\pm$ 0.72 & & 177 $\pm$ 43 & 2.33 $\pm$ 0.71 & & 162 $\pm$ 38 & 2.13 $\pm$ 0.63 \\
UGC~06138& C01-C06 &13 (20)& 31 & 75  $\pm$ 51 & 2.02 $\pm$ 1.41 & & 46  $\pm$ 36 & 0.86 $\pm$ 0.70 & &  94 $\pm$ 58 & 1.75 $\pm$ 1.12 \\
UGC~11131&   C10   &19 (25)& 19 & 180 $\pm$ 81 & 1.53 $\pm$ 0.75 & & 173 $\pm$ 67 & 1.46 $\pm$ 0.63 & & 219 $\pm$112&  1.85 $\pm$ 1.01 \\
UGC~11651& C07,C09 &42 (49)& 38 & 199 $\pm$ 57 & 2.72 $\pm$ 0.93 & & 143 $\pm$ 41 & 1.96 $\pm$ 0.66 & & 201 $\pm$ 59 & 2.74 $\pm$ 0.95 \\
\enddata
\tablenotetext{a}{(1) Galaxy ID. (2) Selected control fields as named on Table 2. (3) Number of detected cluster candidates after subtraction of the expected contamination. The total number of GC candidates detected is indicated in brackets. (4) Number of MW globular clusters that would be detected under the same observational conditions. (5) Expected total number of clusters after comparison with the MW system. (6) Specific frequency derived from column 6. (7-8) and (9-10) The same as (5-6) but for a 15\% smaller and 15\% larger galaxy distance respectively.}

\label{tab:SN}
\end{deluxetable*}

\subsection{Color Distribution}

\begin{figure}
\begin{center}
\includegraphics[angle=0,scale=.4]{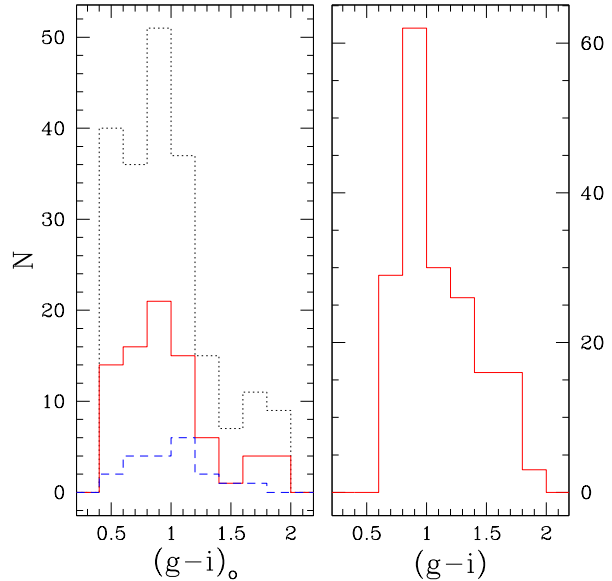}
\caption{Left: The dotted line show the color distribution of globular cluster candidates in the 6 galaxies, compared to the one observed for the 2 intermediate to low surface brightness galaxies ({\it  ugc03587} and {\it ugc06138}) in the solid line, and to the one observed for {\it ugc06138} (dashed line) . Right: Color distribution of the Galactic globular clusters detected under the same observing conditions and selection criteria we applied for the six galaxies.} 
\label{color}
\end{center}
\end{figure}

The left panel of Figure~\ref{color} shows the color distribution of the 206 globular cluster candidates detected in six galaxies as well as the joint color distribution of the two intermediate to low surface brightness galaxies (\object[UGC 03587]{{\it ugc03587}} and \object[UGC 06138]{{\it ugc06138}}) and the individual distribution of \object[UGC 06138]{{\it ugc06138}}. The colors in this panel are not corrected for internal galaxy reddening. The right panel shows the (un-de-reddened) color distribution of the Milky Way GCs that were detected inside the comparison masks of the six galaxies. In other words, the right panel shows how the left panel would look like if the GCs in all the galaxies would have color distributions such as the one of the Milky Way.

Comparing the total sample color distribution to the one corresponding to the Milky Way, we note that there is a higher number of GCs in the bluest bin of our sample, probably due to our relaxed color selection criteria, which can lead to the inclusion of some young star clusters in our sample. The results of a Kolmogorov-Smirnov test performed in order to compare both distributions indicates that the null hypothesis that both samples arise from the same parent distribution cannot be rejected ($p$-value of 0.4) when restricting the color to satisfy $(g-i)>0.68$ mag, a regime where both distributions overlap. We conclude then that the GC candidates observed in the LSB galaxies have a color distribution similar to the one observed in the Milky Way in the region where they overlap. An additional group of objects sharing the general features of GCs but with bluer colors is present in the sample galaxies. Whether this color is due to low metallicities or young age is something we cannot answer with our data but can be addressed with follow-up spectroscopy.

With respect to the color distribution of \object[UGC 03587]{{\it ugc03587}} and \object[UGC 06138]{{\it ugc06138}}, the data suggest that they roughly follow the same distribution as the rest of the galaxies in the sample. While no evident differences are apparent as a function of central surface brightness, we note that we are dealing with small number statistics. 

It has been known for some time that the color distributions of early type galaxies are bimodal (see, e.g.~, Peng et al.~2006b and references therein). This property has also been observed in the GC populations of normal spiral galaxies, including the Milky Way and M31. For the latter galaxies, the color bimodality has been shown to correspond to metallicity bimodality (see e.g., Ashman \& Zepf 1998, Barmby et al.~2000). The number of GC candidates in our galaxies is not high enough to differentiate between uni- or bimodality, but we note that some asymmetry can be observed in the color distributions.

\subsection{Size Distribution}

\begin{figure}
\begin{center}
\includegraphics[angle=0,scale=.7]{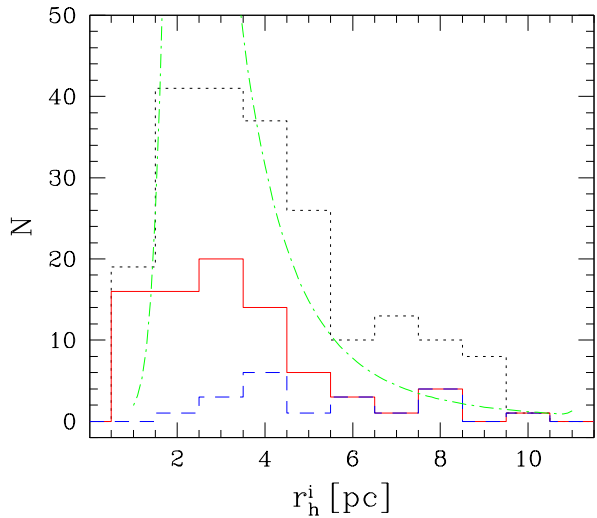}
\caption{Size distribution of globular cluster candidates detected in the 6 galaxies (dotted line) compared to the distribution observed for the 2 intermediate to low surface brightness galaxies ({\it ugc03587} and {\it ugc06138}) in the solid line, and to the one observed for {\it ugc06138} (dashed line). The dot-dashed line shows a fit for the size distribution of GCs in the Milky Way convolved with the error distribution observed in our sample (see text).}
\label{radio}
\end{center}
\end{figure}

The half light radii distribution of GC candidates is displayed in Figure~\ref{radio} for all the galaxies in the sample and for the same sub-samples in surface brightness described in the previous section. We have overplotted in this figure the analytical distribution that Jord\'an et al.~(2005) have shown to be a good fit for the size distribution of GCs in the Milky Way and in early-type galaxies, convolved with the error distribution observed in our sample. We are using their eq. 22 with parameters: $\mu=0.37$, $\beta_1=0.117$, $\beta_2=0.078$ and $f=0.7$. 

Our sample shows a size distribution which is qualitatively similar to the one observed in early-type galaxies and the Milky Way (Jord\'an et al.~2005). In particular, this distribution shows a peak at a half-light radius of $\sim 3$ pc, with an extended tail to larger distances. We note that the fact that the typical sizes of our GC candidates are $\sim 3$ pc is an independent confirmation that our assumed distances are reasonable. As shown in Jord\'an et al.~(2005) following previous suggestions (e.g., Kundu \& Whitmore 2001), the typical sizes of GCs can be used as a standard ruler and therefore biased distances would have returned typical GC sizes in disagreement with the expected values.

In the small half light radius regime we observe a group of very compact objects ($r_{h}<$1.5) that do not seem to fit properly the Galactic size distribution. This population overlaps with the one responsible in creating an excess at the blue region of the color distribution, which indicates that they are most likely internal contamination due to young compact clusters in the disk of the galaxies, whose color and size distributions happen to overlap the ones for globular clusters in these regions (Larsen 1999, Scheepmaker et al.~2007).

\subsection{Luminosity Function}

\begin{figure}
\begin{center}
\includegraphics[angle=0,scale=.4]{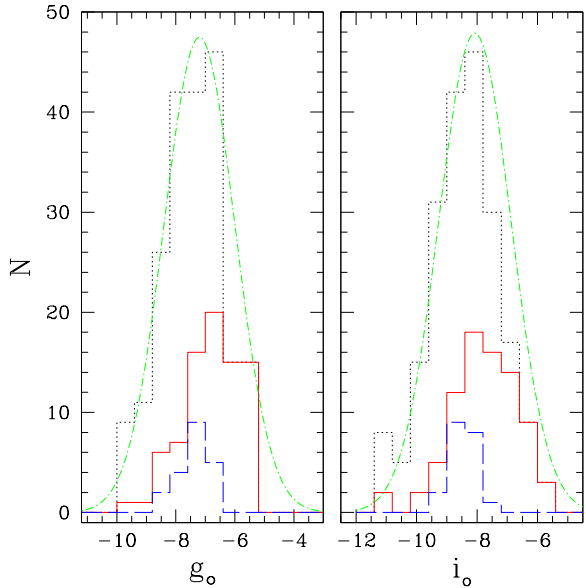}
\caption{Luminosity distribution of the globular cluster candidates following the same line-style used in the previous 2 figures. The dot-dashed line shows the expected GCLF for a galaxy with $M_{B} \sim 19$, a Gaussian distribution with parameters ($\bar{M}_{g}$, $\sigma_{g}$)=(-7.2, 1.2) and ($\bar{M}_{i}$, $\sigma_{i}$)=(-8.1, 1.2) (Jord\'an et al.~2007b).} 
\label{fig:gclf}
\end{center}
\end{figure}

Figure \ref{fig:gclf} shows the luminosity distribution of the globular cluster candidates in all 6 galaxies in both bands. This figure also shows the luminosity distribution of the sub-sample of intermediate to low luminosity galaxies and the individual distribution of \object[UGC 06138]{{\it ugc06138}}. The Gaussian distribution over-plotted in both panels of Figure \ref{fig:gclf} shows the expected globular cluster luminosity function for a galaxy of $M_B$ $\sim$ -19, whose parameters correspond to ($\bar{M}_{g}$, $\sigma_{g}$)=(-7.2, 1.2) and ($\bar{M}_{i}$, $\sigma_{i}$)=(-8.1, 1.2), according to Jord\'an et al.~(2007b). These distributions were normalized to twice the number of cluster candidates brighter than the peak (turnover) of the luminosity function. By normalizing in this fashion we circumvent the effects of incompleteness present at magnitudes fainter than the turnover.

The luminosity distribution of all the cluster candidates gives the appearance of being brighter than expected in both bands. This is due to different foreground absorption conditions between the sample galaxies, which do not allow to reach equally deep magnitudes for all of them. In fact, \object[UGC 03587]{{\it ugc03587}} is the only case where the sample extends $\sim$2.5$\sigma$ down the expected peak of the distribution. The samples of  \object[UGC 00477]{{\it ugc00477}}, \object[UGC 06138]{{\it ugc06138}} and \object[UGC 11651]{{\it ugc11651}} reach at least the turnover, and in the cases of \object[UGC 03459]{{\it ugc03459}} and \object[UGC 11131]{{\it ugc11131}} we are sampling just the most luminous objects. Considering this caveat, the observed distributions are consistent with the expectations from the luminosity function of the MW and early-type galaxies (e.g. Jord\'an et al.~2007b).

It is also worth noticing that the luminosity distributions of clusters hosted by  \object[UGC 03587]{{\it ugc03587}} and \object[UGC 06138]{{\it ugc06138}} are consistent with those of the GCs hosted by high surface brightness galaxies in our sample. This implies, under the standard picture of the dynamical formation of the GC luminosity function (e.g., Fall \& Zhang 2001), that two-body relaxation and evaporation have acted with similar effectiveness in both kind of systems. Given that the evaporation rate can be argued to depend mainly on the GC half-mass density (e.g., McLaughlin \& Fall 2007) we expect the extent of GC dynamical evolution to be similar across our sample given that the luminosity and size distributions show no systematic variations, implying the same for the GC half-mass densities.

\section{Summary and Conclusions}
\label{sec:CONC}
We have studied the GC systems of a sample of six massive late-type galaxies with a range of surface brightness, including two galaxies with blue central disk surface brightness $\mu_{0,\rm disc}>$ 22 mag/arcsec$^{2}$ . 

Through direct comparison with the Milky Way, we estimated the total number of globular clusters and using these values we have derived specific frequencies for each galaxy. The results obtained do not show significant differences between galaxies in different ranges of surface brightness and also indicate that all of them have formed globular clusters with similar efficiency in the range expected for late-type galaxies. 

A general analysis of the properties of the 206 clusters candidates detected in the 6 sample galaxies has shown that they all show similar characteristics as those observed in previously studied GC systems. Their luminosity function turns over close to the expected value of $M_{g}\sim-7.2$ and their size distribution peaks at $\sim$3~pc, just like in the case of the Milky Way and many other spiral and elliptical galaxies. Their color distribution is similar to the one expected if the Galactic globular clusters were observed under the same conditions as the studied galaxies. All these properties do not show significant evolution with the surface brightness of the host galaxy.

When the properties of two intermediate-to-low surface brightness galaxies are compared to the high surface brightness galaxies in the sample they do not show particular features. In all the properties we probed the globular cluster candidates in  \object[UGC 03587]{{\it ugc03587}} and  \object[UGC 06138]{{\it ugc06138}} show no significant differences with respect to the other galaxies even though they have a much lower central surface brightness. This implies that the physical conditions present during the formation of GCs in massive LSB galaxies were not significantly different to those of HSB galaxies.

Our results show that the GC formation process in LSB galaxies proceeded in a way that, at least with the available information, is indistinguishable from the processes observed in normal (i.e high surface) brightness galaxies. Our data suggests a close similarity in the early assembly processes of both types of galaxies, which implies that despite a very quiescent star formation history thereafter, the initial period of assembly must have been intense enough to allow the formation of massive star clusters. Improved constraints can be obtained after studying the spectra of the globular clusters in order to date their formation time and measure the metal abundances of the environment in which they where formed.

We have found the presence of some ``contamination'' in the sample of cluster candidates, coming from a group of blue and small ($r_{h}<$1.5) objects. We are unable to determine whether the blue color is due to low metallicity or young ages. The first possibility would mean that these galaxies were able to form lower metallicity objects than the ones observed in the Milky Way, while the second one would imply that ongoing clustered star formation in the observed galaxies (and particularly in  \object[UGC 03587]{{\it ugc03587}} which is the main contributor to this population) is driven in a fashion similar to that observed in normal brightness galaxies, despite the low-star formation rates expected based on the Kennicutt law (Kennicutt 1989).

It is worth mentioning that all the 6 galaxies included in our sample were originally cataloged as ``low surface brightness galaxies'', but after a detailed decomposition of their surface brightness profiles we have found just two of them to be consistent with this classification. This stresses the importance of performing a surface brightness decomposition and proper absorption corrections when classifying objects as LSB galaxies.

The results we have presented here are broadly consistent with the scenario proposed by van den Hoek et al.~(2000) in which LSB galaxies roughly follow the same evolutionary history as HSB galaxies except at a much lower rate. We have show that the early formation process must have been intense enough in both types of galaxies to allow the formation of massive star clusters, even though the ensuing evolution proceeds at very different rates.

\acknowledgments

Support for program GO-10550 was provided through a grant from the Space
Telescope Science Institute, which is operated by the Association of 
Universities for Research in Astronomy, Inc., under NASA contract NAS5-26555. 
This research has made use of the NASA/IPAC Extragalactic Database (NED)
which is operated by the Jet Propulsion Laboratory, California Institute
of Technology, under contract with the National Aeronautics and Space 
Administration. 

{\it Facility:} \facility{HST (ACS/WFC)}

\clearpage
\appendix 
\section{Notes on individual galaxies}

\begin{itemize}
\item \object[UGC 00477]{{\it ugc00477}} and \object[UGC 06138]{{\it ugc06138}}. These are the two galaxies for which we used a wide selection of control fields to estimate the contamination by background galaxies, due to their location at high Galactic latitude, which also contributes to minimize the uncertainty due to reddening conditions. The distance to the galaxies makes prohibitive a spectroscopic follow up with 8-m class telescopes.
 
\item \object[UGC 03459]{{\it ugc03459}}. This is the galaxy with the most uncertain results. 
Its field of view is located directly through the galactic plane, 
at a latitude of $b=-3.92$, where any local reddening feature will 
produce dramatic changes in the results. The almost null detection of candidates in this galaxy tells us 
that the objects are too absorbed to be detected.

\item \object[UGC 03587]{{\it ugc03587}} and \object[UGC 11651]{{\it ugc11651}}. These galaxies have  a  high number of GC candidates detected. Nevertheless, some of the candidates are projected along the spiral arms of the galaxies, so some contamination due to young star clusters is expected. We also noticed that due to the very peculiar morphology of \object[UGC 03587]{{\it ugc03587}}, SExtractor's model used to subtract the light of this galaxy when detecting point sources is not as good as in the other galaxies. These galaxies are both close enough to be followed up with ground-based spectroscopy in 8-m class telescopes.

\item \object[UGC 11131]\object[UGC 11131]{{\it ugc11131}}. Due to its location close to the direction of the Galactic center this galaxy is affected by strong extinction. It is also very difficult to obtain good control fields in its case as in this region the stellar density has a strong dependence on the galactic longitud, thus this is the only galaxy with just one control field.

\end{itemize}

\begin{thebibliography}{} 

\bibitem[Ashman(1998)]{ashman98} Ashman, K.M. \& Zepf, S.E., 1998, Globular Clusters Systems (Cambridge Univ. Press) 
\bibitem[Barmby(2000)]{barmby00} Barmby, P., Huchra, J.P., Brodie, J.P., Forbes, D.A., Schroder, L.L., \& Grillmair, C.J. 2000, AJ, 119, 727
\bibitem[Bertin_Arnouts(1996)]{SExtractor} Bertin, E. \& Arnouts, S. 1996, \aaps, 117, 393
\bibitem[Boissier(2003)]{boissier03} Boissier, S., Monnier Ragaigne, D., Prantzos, N., van Driel, W., Balkowski, C. \& O'Neil, K., 2003, \mnras, 343, 653 
\bibitem[Bothun(1985)]{bothun85} Bothun, G.D., Beers, T.C., Mould, J.R. \& Huchra, J.P., 1985, \aj, 90, 2487
\bibitem[Bothun(2002)]{bothun02} Bothun, G.D., Schombert, J.M., Impey, C.D., Sprayberry, D. \& McGaugh, S.S., 
1993, \aj, 106, 530
\bibitem[Byun(1996)]{byun96} Byun, Y.I., Grillmair, C.J., Faber, S.M., Ajhar, E.A., Dressler, A., Kormendy, J., Lauer, T.R., Richstone, D.\& Tremaine, S., 1996, AJ, 111, 1889
\bibitem[Churchwell(2002)]{church02} Churchwell, E., 2002, \araa 40, 27
\bibitem[Cote(2004)]{acsvcs_i} C\^ot\'e, P., Blakeslee, J.P., Ferrarese, L., Jord\'an, A., Mei, S., Merritt, D., Milosavljevic, M., Peng, E.W., Tonry, J.L., \& West, M.J., 2004, \apjs, 153, 223 
\bibitem[Dalcanton(1997)]{dalcanton97} Dalcanton, J.J., Spergel, D.N. \& Summers, F.J., 1997, \apj, 482, 659
\bibitem[Davis(1997)]{davis97} Davis, M., Miller, A \& White S.D.M., 1997,\apj, 490, 63
\bibitem[Disney(1976)]{disney76} Disney, M., 1976, Nature, 263, 573
\bibitem[deBlok(1995)]{deBlok95} de Blok, W.J.G., van der Hulst, J.M. \& Bothun, G.D., 1995, \mnras, 274, 235
\bibitem[deBlok(1998)]{deBlok98} de Blok, W.J.G., van der Hulst, J.M., 1998, A\&A, 335, 421
\bibitem[Ferrarese(2006)]{Ferrarese06} Ferrarese, L., C\^ot\'e, P., Jord\'an, A., Peng, E.W., Blakeslee, J.P., Piatek, S., Mei, S., Merritt, D., Milosavljevic, M., Tonry, J.L., \& West, M.J. 2006, \apjs, 164, 334
\bibitem[Freeman(1970)]{Freeman70} Freeman, K.C., 1970, \apj, 160,811
\bibitem[Fukugita(1995)]{fukugita} Fukugita, M., Shimasaku, K. \& Ichikawa, T. 1995, PASP, 107, 945 
\bibitem[Galaz2006]{galaz06} Galaz, G., Villalobos, A., Infante, L. \& Donzelli, C., 2006, \aj, 131, 2035
\bibitem[Graham(1998)]{graham05} Graham, A.W. \& Driver, S.P., 2005, Publ. Astron. Soc. Australia, 22, 118
\bibitem[Harris(1996)]{harris81} Harris, W. E. \& van den Bergh, S., 1981, \aj, 429, 177
\bibitem[Harris(1991)]{harris91} Harris, W.E., 1991, \araa, 29, 543
\bibitem[Harris(1996)]{harris96} Harris, W.E., 1996, \aj, 112, 1487
\bibitem[Harris(2001)]{harris01} Harris, W.E., 2001, in Star Clusters, ed. L. Labhardt \& B. Binggeli (Berlin: Springer), 223 
\bibitem[Impey \& Bothun 1996]{impey_bothun96} Impey, C., \& Bothun, G. 1997, \araa, 35, 267
\bibitem[Jedrzejewski 1987]{Jedrzejewski87} Jedrzejewski, R. I., 1987, \mnras, 226, 747
\bibitem[Jimenez et al.~1998]{jimenez+98} Jimenez, R., Padoan, P., Matteucci, F., \& Heavens, A. 1998, \mnras, 299, 123
\bibitem[Jordan(2004)]{jordan04} Jord\'an, A., Blakeslee, J.P., Peng, E.W., Mei, S., Cote, P., Ferrarese, L., Tonry, J.L., Merritt, D., Milosavljevic, M., \& West, M.J. 2004, \apjs, 154, 509
\bibitem[Jordan(2005)]{jordan05} Jord\'an, A., C\^ot\'e, P., Blakeslee, J.P., Ferrarese, L., McLaughlin, D., Mei, S., Peng, E.W., Tonry, J.L., Merritt, D., Milosavljevic, M., Sarazin, C.L., Sivakoff, \& West, M. 2005, \apj, 634, 1002
\bibitem[Jordan(2006)]{gclf_lett} Jord\'an, A., McLaughlin, D.E., C\^ot\'e, P., Ferrarese, L., Peng, E.W., Blakeslee, J.P., Mei. S., Villegas, D., Merritt, D., Tonry, J.L., \& West, M.J. 2006, \apj, 651, L25.
\bibitem[Jordan(2007a)]{acsfcs_i} Jord\'an, A., Blakeslee, J.P., C\^ot\'e, P., Ferrarese, L., Infante, L., Mei, S., Merritt, D., Peng, E.W., Tonry, J.L. \& West, M.J., 2007a, \apjs, 169, 213
\bibitem[Jordan(2007b)]{gclf} Jord\'an, A., McLaughlin, D.E., C\^ot\'e, P., Ferrarese, L., Peng, E.W., Mei. S., Villegas, D., Merritt, D., Tonry, J.L., \& West, M.J. 2007b, \apjs, in press.
\bibitem[Kaisler(1996)]{kaisler96} Kaisler, D., Harris, W.E., Crabtree, D.R. \& Richer, H.B., 1996, \aj, 111, 2224
\bibitem[Kennicutt(1989)]{kennicutt89} Kennicutt, R.C., 1989, \apj, 344, 685 
\bibitem[King(1966)]{king66} King, I.R., 1966, \aj, 71, 64
\bibitem[Kissler(1999)]{KP99} Kissler-Patig, M., Ashman, K.M., Zepf, S.E. \& Freeman, K.C., 1999, \aj, 118, 197
\bibitem[Koekemoer(2001)]{koekemoer01} Koekemoer, A.M., Fruchter, A.S., Hook, R.N. \& Hack, W., 2002, in The 2002 HST Calibration Workshop. Ed. S. Arribas, A.Koekemoer \& B. Whitmore, (Baltimore, STScI), 337
\bibitem[Kundu(2001)] Kundu, A. \& Whitmore, B.C., 2001, 121, 2950
\bibitem[Larsen(1999)]{Larsen99} Larsen, S.S. 1999, A\&AS, 139, 393
\bibitem[McLaughlin(1999)]{mclaughlin99} McLaughlin, D.E., 1999, \aj, 117, 2398
\bibitem[Maraston(1998)]{maraston98} Maraston, C., 1998, \mnras, 300, 872
\bibitem[Maraston(2005)]{maraston05} Maraston, C., 2005, \mnras, 362, 799
\bibitem[McGaugh(1995a)]{mcgaugh95a} McGaugh, S.S., Schombert, J.M. \& Bothun, G.D., 1995, \aj, 109, 2019
\bibitem[McGaugh(1995b)]{mcgaugh95b} McGaugh, S.S., Bothun, G.D. \& Schombert, J.M., 1995, \aj, 110, 573
\bibitem[Mo(1994)]{mo94} Mo, H.J., McGaugh, S.S. \& Bothun, G.D., 1994, \mnras, 267, 129
\bibitem[Mould(2000)]{mould00} Mould, J.R., Huchra, J.P., Freedman, W.L., Kennicutt, R.C., Ferrarese, L., Ford, H.C., Gibson, B.K., Graham, J.A., Hughes, S.M.G., Illingworth, G.D., Kelson, D.D., Macri, L.M., Madore, B.F., Sakai, S., Sebo, K.M., Silbermann, N.A. \& Stetson, P.B., 2000, \apj, 529, 786 
\bibitem[Nilson(1973)]{nilson73} Nilson, P., 1973, Uppsala General Catalogue of Galaxies, Uppsala Astr. Obs. Ann., v.6
\bibitem[Nelder(1965)]{nelder65} Nelder, J.A., \& Mead, R. 1965, Comput. J., 7, 308
\bibitem[ONeil(2004)]{oneil04}  O'Neil, K., Bothun, G., van Driel, W. \& Monnier Ragaigne, D., 2004, A\&A, 428, 823
\bibitem[Peng(2006a)]{peng06a} Peng, E.W., Jord\'an, A., C\^ot\'e, P., Blakeslee, J.P., Ferrarese, L., Mei, S., West, M.J., Merritt, D., Milosavljevic, M., \& Tonry, J.L., 2006a, \apj, 639, 95
\bibitem[Peng(2006b)]{peng06b} Peng, E.W., C\^ot\'e, P., Jord\'an, A., Blakeslee, J.P., Ferrarese, L., Mei, S., West, M.J., Merritt, D., Milosavljevic, M. \& Tonry, J.T., 2006b, \apj, 639, 838
\bibitem[Rosenbaum(2004)]{rosen04} Rosenbaum, S.D. \& Bomans, D.J., 2004, A\&A, 422
\bibitem[Scheepmaker(2007)]{scheep06} Scheepmaker, R.A., Haas, M.R., Gieles, M., Bastian, N., Larsen, S.S. \& Lamers, H.J.G.L.M., 2007, \aap in press (arXiv:0704.3604)
\bibitem[Schombert(1990)]{schombert90} Schombert, J.M., Bothun, G.D., Impey, C.D. \& Mundy L.G., 1990, \aj, 100, 1523 
\bibitem[Schombert(1992)]{schombert92} Schombert, J.M., Bothun, G.D., Schneider, S.E. \& McGaugh, S.S., 1992, \aj, 103, 1107
\bibitem[Schlegel(1998)]{schlegel98} Schlegel, D.J., Finkbeiner, D.P. \& Davis, M., 1998, \apj, 500, 525
\bibitem[Sharina(2005)]{sharina05} Sharina, M.E., Puzia, T.H. \& Makarov, D.I., 2005, A\&A, 442, 85 
\bibitem[Sirianni(2005)]{sirianni05} Sirianni, M., et al.~2005, \pasp, 117, 1049S
\bibitem[vandenhoek(1993)]{hoek00} van den Hoek, L.B., de Blok, W.J.G., van der Hulst, J.M. \& de Jong, T., 2000, A\&A, 357, 397
\bibitem[vanderhulst(1993)]{hulst93} van der Hulst, J.M., Skillman, E.D., Smith, T.R., Bothun, G.D., McGaugh, S.S. \& de Blok, W.J.G., 1993, \aj, 106, 548
\bibitem[West(2004)]{west+04} West, M.J., C\^ot\'e, P., Marzke, R.O., \& Jord\'an, A. 2004, Nature, 
472, 31
\bibitem[White(1987)]{white_shawl_87} White, R.E., \& Shawl, S.J. 1987, \apj, 317, 246
\bibitem[Zackrisson(2005)]{zack05} Zackrisson, E., Bergvall, N., \& \"Ostlin, G., 2005, A\&A, 435, 29
\bibitem[Zinn(1985)]{zinn85} Zinn, R., 1985, \apj, 293, 424
\bibitem[Zwann et al.~(1995)]{zwann95} Zwaan, M.A., van der Hulst, J.M., de Blok, W.J.G., \& McGaugh, S.S. 1995, \mnras, 273, L35

\end{thebibliography}
\end{document}